\documentclass[superscriptaddress,twocolumn]{revtex4-1}
\pdfoutput=1
\usepackage[colorlinks=true,urlcolor=blue]{hyperref}
\usepackage{amsfonts}
\usepackage{graphicx}
\usepackage{amsmath}
\usepackage{xcolor}
\usepackage{bm}

\usepackage{color}
\definecolor{red}{rgb}{0.75,0,0}
\definecolor{blue}{rgb}{0,0,0.75}
\definecolor{green}{rgb}{0,0.5,0}

\newcommand{\revision}[1]{#1}

\DeclareMathOperator{\tr}{tr}
\DeclareMathOperator{\sgn}{sgn}

\begin{document}

\title{Orientational properties of nematic disclinations}

\author{Arthur J. Vromans}
\affiliation{Instituut-Lorentz, Universiteit Leiden, P.O. Box 9506, 2300 RA Leiden, The Netherlands}
\author{Luca Giomi}
\affiliation{Instituut-Lorentz, Universiteit Leiden, P.O. Box 9506, 2300 RA Leiden, The Netherlands}

\begin{abstract}
Topological defects play a pivotal role in the physics of liquid crystals and represent one of the most prominent and well studied aspects of mesophases. While in two-dimensional nematics, disclinations are traditionally treated as point-like objects, recent experimental studies on active nematics have suggested that half-strength disclinations might in fact possess a polar structure. In this article, we provide a precise definition of polarity for half-strength nematic disclinations, we introduce a simple and robust method to calculate this quantity from experimental and numerical data and we investigate how the orientational properties of half-strength disclinations affect their relaxational dynamics.
	
\end{abstract}

\maketitle

\section{Introduction}

Disclinations are singularities in nematic liquid crystals where the average molecular orientation is undefined \cite{DeGennes:1993,Chaikin:1995,Kleman:2003}. These defects are ubiquitous in nematic samples and represent their most prominent and visible feature \cite{Musevic:2006,Ravnik:2007,Ravnik:2009,Copar:2012,Senyuk:2013,Copar:2013,Martinez:2014}. In fact, the existence itself of a nematic phase, where the molecules are oriented along a common direction, but have zero macroscopic polarization, was established only once disclinations were correctly identified \cite{Frank:1958}. 

In two-dimensional nematic liquid crystals, disclinations are conventionally treated as point-like objects interacting with each other through elastic forces formally analogous to the Coulomb force in electrostatics \cite{DeGennes:1993,Chaikin:1995,Kleman:2003,Chandrasekhar:1986,Rosso:2001,Toth:2002,deLozar:2005,Alexander:2012}. Recently, however, combined experimental and theoretical efforts toward understanding the mechanics of active nematics (i.e. nematic liquid crystals obtained from self- or mutually-propelled rod-like macromolecules, typically of biological origin), have suggested that some type of nematic disclinations have in fact a polar structure and can reorient each other via elastic and hydrodynamic torques \cite{Keber:2014,DeCamp:2015,Doostmohammadi:2015,Oza:2015}. Two spectacular examples of these phenomena have been recently reported in two-dimensional active nematic suspensions of microtubule bundles and kinesin. Keber {\em et al}. \cite{Keber:2014} constructed an active nematic vesicle by encapsulating microtubules, kinesin motors clusters and polyethylene glycol (PEG) within a lipid vesicle. A depletion mechanism, due to the PEG, drives the microtubules to the inner leaflet of the vesicle, giving rise to a dense two-dimensional nematic cortex. As a consequence of the spherical confinement, such a two-dimensional nematic is forced to contain four $+1/2$ disclinations \cite{Nelson:2002,Vitelli:2006,LopezLeon:2011}, which, due to the local hydrodynamic flow fueled by the active stresses, travel at constant speed toward the ``head'' of their comet-like structure \cite{Keber:2014}. More recently, DeCamp {\em et al.} \cite{DeCamp:2015} demonstrated that defects in two-dimensional active nematics can themselves form a nematic phase in which the head-to-tail directionality of $+1/2$ disclinations propagates over distances several order of magnitude larger than the length of a single microtubules \cite{DeCamp:2015}. 

In spite of this convincing experimental evidence, a precise definition of defect orientation is, however, still lacking. \revision{Intuitively, the existence of an orientation {\em per se} does not depend on whether the system is passive or active, but it is only determined by the defects local geometry. Here we consider the simplest possible setting, consisting of a two-dimensional passive nematic liquid crystal subject to a purely relational dynamics, and we address the following fundamental questions:} what is behind the comet-like appearance of $+1/2$ disclinations? Can an orientation be defined for $-1/2$ disclinations as well? How can the orientation of a defect be determined from numerical and experimental data and, perhaps more importantly, does such an orientation affect the mechanics of half-strength disclinations in any way? 
 
We start by providing a rigorous definition of defect orientation and introduce a simple method to calculate the orientation of half-strength disclinations from pixelated data. Next, using numerical simulations, we demonstrate that pairs of like-sign disclinations exert elastic torques that tends to anti-align them as they repel, while oppositely charged disclinations have negligible orientational interaction. For both cases, however, the coupling between translational and rotational dynamics is very strong and leads to a variety of novel annihilation/repulsion trajectories. Finally, we analytically address the ideal case of two like-sign disclinations on an infinite plane and show that these interact through an elastic torque independent on their distance and reminiscent of that between dislocations in two-dimensional solids \cite{Nabarro:1987,Eisenmann:2005}. As a consequence, pairs of like-sign disclinations on an infinite plane anti-align exponentially with time. 

\section{Results}

\subsection{Polarity of half-strength nematic disclinations}

\begin{figure}[t]
\centering
\includegraphics[width=1\columnwidth]{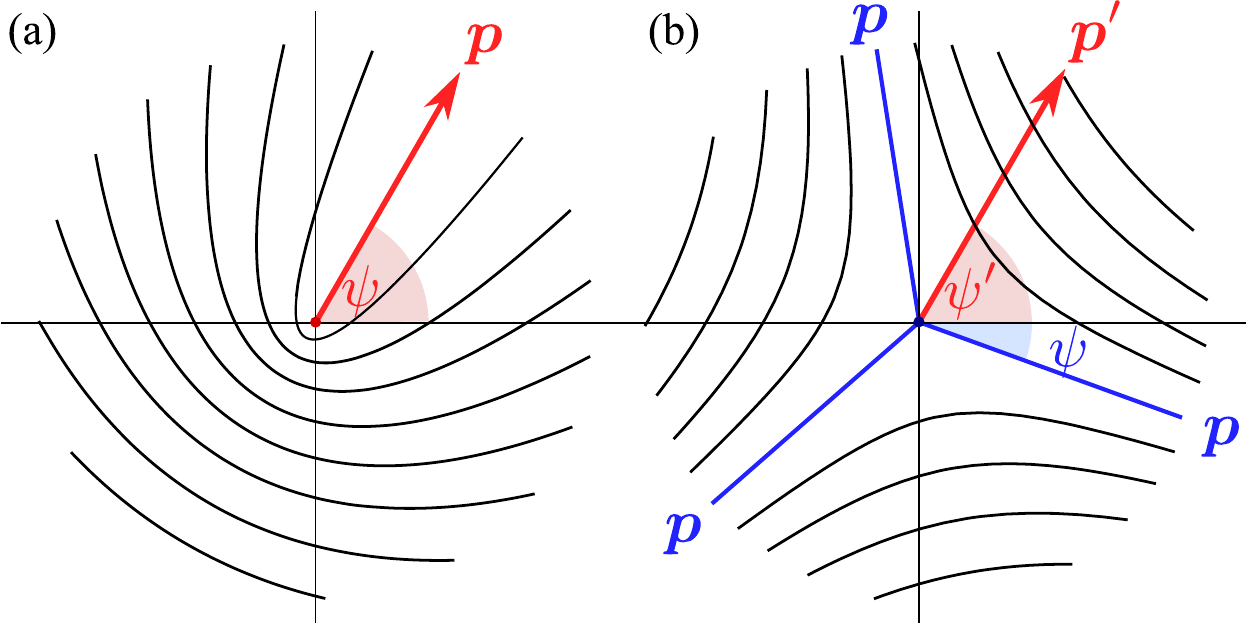}  
\caption{\label{fig:snapshots}Examples of $+1/2$ (a) and $-1/2$ (b) disclinations with generic orientation $\bm{p}=(\cos\psi,\sin\psi)$, with $\psi$ calculated from Eq. \eqref{eq:divergence}. For $-1/2$ disclinations, $\bm{p}$ is related to the polarity $\bm{p}'=(\cos\psi',\sin\psi')$ of the dual $+1/2$ disclination obtained from the mirror-reflection $\theta\rightarrow-\theta$, as $\psi=-\psi'/3$.}
\end{figure}

Let us consider a nematic liquid crystal in two dimensions and let $\bm{n}=(\cos\theta,\sin\theta)$ be the nematic director representing the average molecular orientation. In the presence of disclinations, the director rotates by a multiple of $\pi$ in one loop around the defect core, thus: $\oint d\theta=2\pi k$, where $k=\pm 1/2,\, \pm 1\,\ldots$ is the turning number of ``strength'' of the disclination and the contour integral is calculated on any closed path enclosing the core. In one elastic constant approximation, a well-known defective solution that minimizes the Frank energy:
\begin{equation}
E_{\rm F} = \frac{1}{2}K\int dA\,|\nabla\theta|^{2}\;,
\end{equation}
with $K$ an elastic constant, is given by \cite{DeGennes:1993}: 
\begin{equation}\label{eq:theta_0}
\theta=k\phi+\theta_{0}\;,
\end{equation}
where $\phi=\arctan(y/x)$ is the usual polar angle. The constant $\theta_{0}$, which is often omitted, describes a global rotation of the director about the $z-$axis passing through the defect core. Now, as it is evident from Fig. \ref{fig:snapshots}, these energy minimizing defective configurations have $2|1-k|-$fold rotational symmetry. This can be made explicit by introducing a new constant angle $\psi$, such that $\theta_{0}=(1-k)\psi$. Eq. \eqref{eq:theta_0} becomes then:
\begin{equation}\label{eq:theta}
\theta = k(\phi-\psi)+\psi\;. 	
\end{equation} 
For $k=1/2$, $\theta=(\phi+\psi)/2$ and it takes a full $2\pi$ rotation in order to transform $\bm{n}$ into itself: i.e. $\bm{n}(\psi)=\bm{n}(\psi+2m\pi)$, with $m$ an integer. While for $k=-1/2$, $\theta=-(\phi-3\psi)/2$ and $\bm{n}$ is manifestly invariant under rotations by multiples of $2\pi/3$: i.e. $\bm{n}(\psi)=\bm{n}(\psi+2m\pi/3)$. The vector: 
\begin{equation}\label{eq:p}
\bm{p}=(\cos\psi,\sin\psi)\;,
\end{equation}
with $|\psi| \le \pi/[2(1-k)]$ defines then the polarity of disclinations of strength $k=\pm 1/2$ (see Fig. \ref{fig:snapshots}). By construction, $\bm{p}$ is defined up to rotations by $\pi/(1-k)$. 

We now turn to the question of calculating the defect orientation from an arbitrary defective configuration of the nematic director. Both in experiments and numerical simulations, local orientations are generally available in the from of a discrete set of data points on a grid, say $\{\theta_{i}\}$, with $i$ the label of a generic grid point. Disclinations can then be detected by measuring the turning number of each plaquette $\mathcal{P}$ in the grid \cite{Huterer:2005}. Then, the simplest method to track the orientation consists of constructing a sequence of data pairs $(\theta_{i},\phi_{i})$, with $i\in\mathcal{P}$ and $\phi_{i}$ measured with respect to the center of the plaquette, and using directly Eq. \eqref{eq:theta} to find $\psi$ through a linear regression. This method, which was adopted in Ref. \cite{DeCamp:2015}, is however often unsatisfactory. Eq. \eqref{eq:theta} corresponds indeed to the ideal, energy minimizing, configuration of a disclination, while in practice the nematic director will always appear distorted. As a consequence, the data points $(\theta_{i},\phi_{i})$ will never align along a straight line, but rather form an undulated curve (see the supplementary information of Ref. \cite{DeCamp:2015}), thus introducing the risk of uncontrolled systematic errors. 

In order to overcome this difficulty, we introduce an alternative method directly inspired to the mechanics of disclinations in active nematics \cite{Keber:2014,Giomi:2014}. In Ref. \cite{Giomi:2014} it was demonstrated that active $+1/2$ disclinations self-propel in the direction of their symmetry axis by virtue of the spontaneous flow powered by the active stresses. The body force driving the active flow is: $\bm{f}^{\rm a}\propto\nabla\cdot{\bm{Q}}=\bm{p}/(2r)$  with $Q_{ij}=S(n_{i}n_{j}-\delta_{ij}/2)$ the nematic tensor \cite{DeGennes:1993}, $S$ the order parameter and $r$ the distance from the defect core. The polarity of $+1/2$ disclinations can then be simply calculated from the divergence of the nematic tensor: $\bm{p}=\nabla\cdot\bm{Q}/|\nabla\cdot\bm{Q}|$. In order to extend this method to $-1/2$ disclinations, we notice that $+1/2$ and $-1/2$ disclinations can be transformed into one another by a mirror reflection of the director: $\theta\rightarrow -\theta$. The polarity $\bm{p}'=(\cos\psi',\sin\psi')$ of the dual $+1/2$ disclination, obtained by mirror-reflecting a $-1/2$ disclination, is straightforwardly related to that of the original $-1/2$ disclination as $\psi=-\psi'/3$ (Fig. \ref{fig:snapshots}). This allows us to calculate the angle $\psi$ for both positive and negative disclinations:
\begin{equation}\label{eq:divergence}
\psi = \frac{k}{1-k}\arctan\left[\frac{\langle \sgn(k)\,\partial_{x}Q_{xy}-\partial_{y}Q_{xx}\rangle}{\langle\partial_{x}Q_{xx}+\sgn(k)\,\partial_{y}Q_{xy}\rangle}\right]\;,
\end{equation}
where $\langle\cdots\rangle$ denotes an average along the shortest available loop enclosing the core and we used the fact that $Q_{xy}\rightarrow -Q_{xy}$ under mirror reflections, hence the sign function $\sgn(k)$. Eq. \eqref{eq:divergence} provides a simple and robust way to calculate the orientation of a defect that is suitable to be implemented on structured and unstructured grids upon approximating the derivatives by finite differences and using a plaquette to calculate the average. \href{http://wwwhome.lorentz.leidenuniv.nl/~giomi/sup_mat/20150720}{Supplementary Movie S1} shows an example of this algorithm in a system undergoing a fast coarsening from a random configuration.

Evidently, the auxiliary vector $\bm{p}'$, used in the derivation of Eq. \eqref{eq:divergence}, can be also employed to calculate the relative orientation of two $-1/2$ disclinations. It is worth to stress, however, that $\bm{p}'$ is not uniquely defined as its direction depends on the choice of the axis used for the mirror transformation (the $x-$axis in this case). Choosing a different axis, say $\bm{\hat{x}'}=(\cos\alpha,\sin\alpha)$, yields a different vector $\bm{p}'$ with $\psi'=-3(\psi-\alpha)$. This ambiguity, however, has no effect on the relative orientation of an arbitrary number of $-1/2$ disclinations. Furthermore, as the Frank free energy is invariant with respect to mirror reflection, the energy of an arbitrary distribution of $-1/2$ disclinations is equal to that of a distribution of $+1/2$ disclinations having the same positions and relative orientations. The consequence of this invariance will be analyzed in the following sections.

\subsection{Orientational energy}

\begin{figure}[t]
\centering
\includegraphics[width=1\columnwidth]{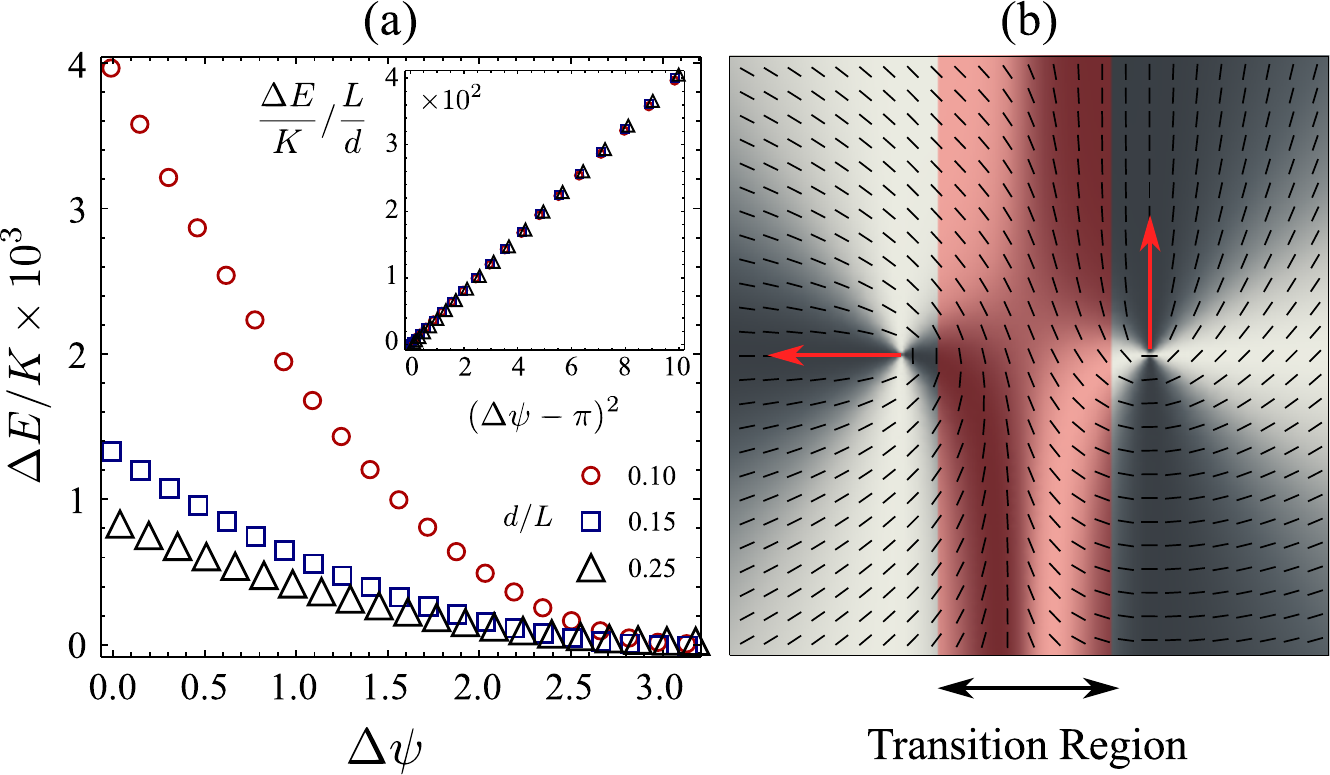}  
\caption{\label{fig:energy}(a) The energy difference $\Delta E = E_{\rm LdG}(\Delta\psi)-E_{\rm LdG}(\pi)$, with $E_{\rm LdG}$ given in Eq. \eqref{eq:landau_de_gennes}, of a pair of $+1/2$ disclinations as a function of their relative tilt $\Delta\psi$ for various distances $d$ of the cores. Inset: the same data rescaled by $L/d$ and plotted versus $(\Delta\psi-\pi)^{2}$, showing a dependence of the form $\Delta E \sim KL/d(\Delta\psi-\pi)^{2}$. (b) Two $+1/2$ disclinations with $\Delta\psi=\pi/2$. The rotation of the director within the \revision{transition region (shaded) between the cores}, determines the angular dependence of the elastic energy.}
\end{figure}

Now that a notion of orientation has been precisely defined and a method to calculate it has been introduced, we can examine how this affects the mechanics of half-strength disclinations. With this goal, we introduce the Landau-De Gennes energy \cite{DeGennes:1993,Olmsted:1992}:
\begin{equation}\label{eq:landau_de_gennes}
E_{\rm LdG} = \frac{1}{2}K\int dA\,\left[|\nabla\bm{Q}|^{2}+\frac{1}{\epsilon^{2}}\tr\bm{Q}^{2}(\tr{\bm{Q}}^{2}-1)\right]\;,
\end{equation} 
with $\epsilon$ a constant with dimensions of length and proportional to the defect core radius. Eq. \eqref{eq:landau_de_gennes} is better suited for numerical applications than the Frank free energy as, unlike the nematic director $\bm{n}$, the $\bm{Q}$ tensor is defined also at the defects (where the order parameter $S$ drops to zero). In order to quantify the energetic cost of defect orientation, we consider two $+1/2$ disclinations in a square $L\times L$ domain, positioned at $\bm{r}_{1}=(-d/2,0)$ and $\bm{r}_{2}=(d/2,0)$ and having $\psi_{1}=\pi$ and varying $\psi_{2}$. Intuitively, we might expect the antiparallel configuration for which $\Delta\psi=\psi_{1}-\psi_{2}=\pi$ to be energetically favorable. Fig. \ref{fig:energy}a shows the difference $\Delta E = E_{\rm LdG}(\Delta\psi)-E_{\rm LdG}(\pi)$ as a function of the angular displacement $\Delta\psi$ and for various distances $d$. The data show a clear dependence of the form $\Delta E \sim K L/d\,(\Delta\psi-\pi)^{2}$ (inset). 

The origin of this behavior is not difficult to understand and provides important insights in the orientational mechanics of nematic disclinations. A pair of $+1/2$ disclinations consists of three regions: the regions surrounding the cores where $\theta=\arctan[(y-y_{i})/(x-x_{i})]/2$ with $i=1,\,2$, and \revision{an intermediate transition region} (Fig. \ref{fig:energy}b). Within the \revision{transition region}, the director is forced to rotate by an amount proportional to $\pi-\Delta\psi$ in order for $\theta$ to match the orientation of the core regions. If the defects are separated by a distance $d$, the area of the \revision{transition region} is roughly $A_{\rm b}\sim Ld$ while the energy density scales like $e_{\rm b}/K \sim (\Delta\psi-\pi)^2/d^2$. Thus the total energy introduced by the \revision{transition region} is $e_{\rm b}A_{\rm b}\sim (KL/d)(\Delta\psi-\pi)^{2}$. \revision{In the general case, the energy of a pair of like-sign defects will depend on the detailed geometry of the transition region (i.e. how smoothly or abruptly the director rotates while interpolating between the cores), but the energy scaling will remain unchanged.}

\begin{figure}[t]
\centering
\includegraphics[width=1\columnwidth]{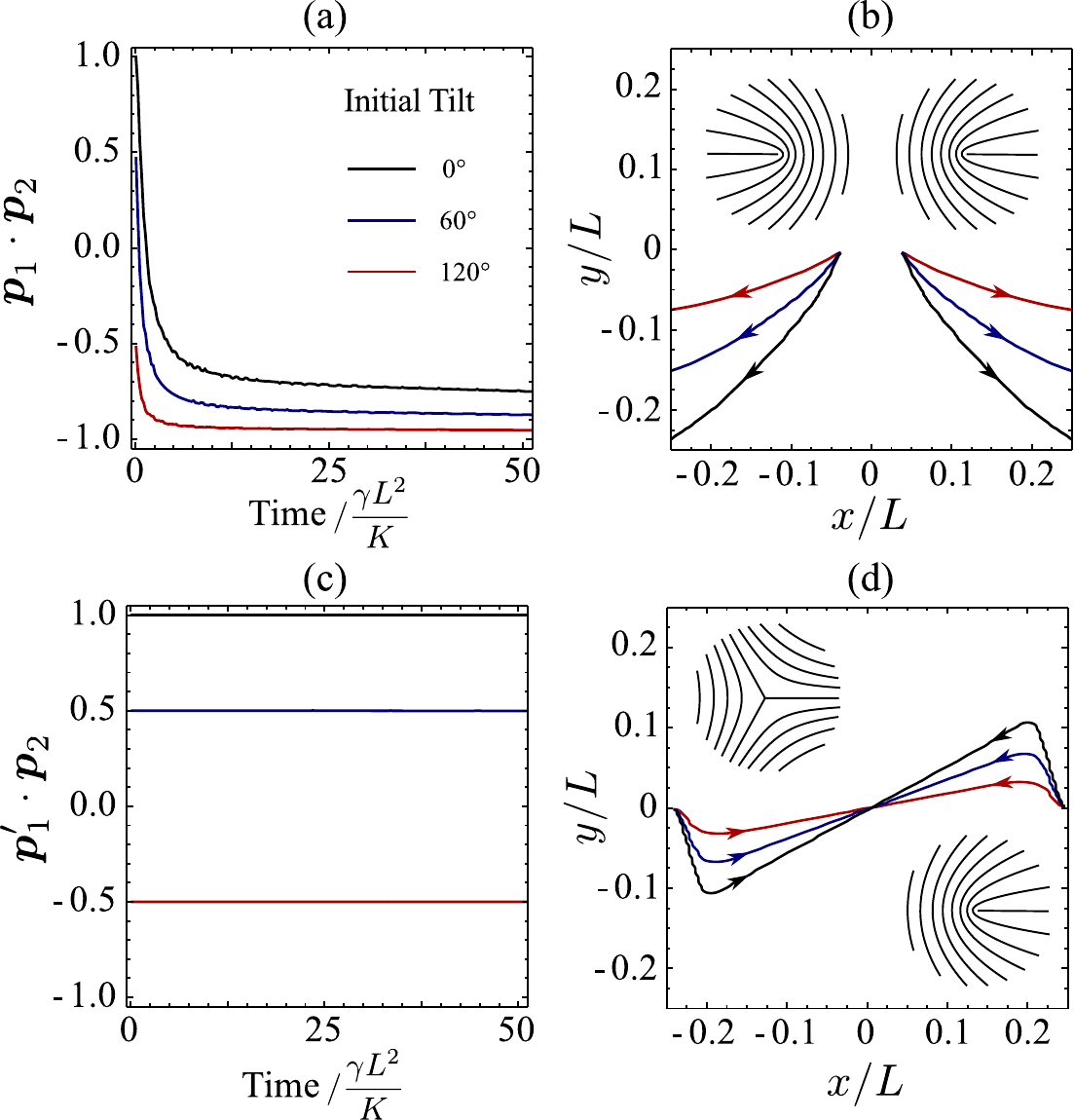}  
\caption{\label{fig:trajectories}The product $\bm{p}_{1}\cdot\bm{p}_{2}$ as a function of time for a pair of $+1/2$ (a) disclinations and the corresponding trajectories (b) in the $xy-$plane. The data are obtained from a numerical integration of Eq. \eqref{eq:dynamics} with $\epsilon/L=5\times 10^{-3}$, initial distance $d/L=0.05$ and various initial tilt values. The product $\bm{p}_{1}'\cdot\bm{p}_{2}$ (c) and the corresponding trajectory (d), for a $\pm 1/2$ pair.}
\end{figure}

\subsection{Relaxational dynamics of disclination pairs}

\begin{figure}[t]
\centering
\includegraphics[width=1\columnwidth]{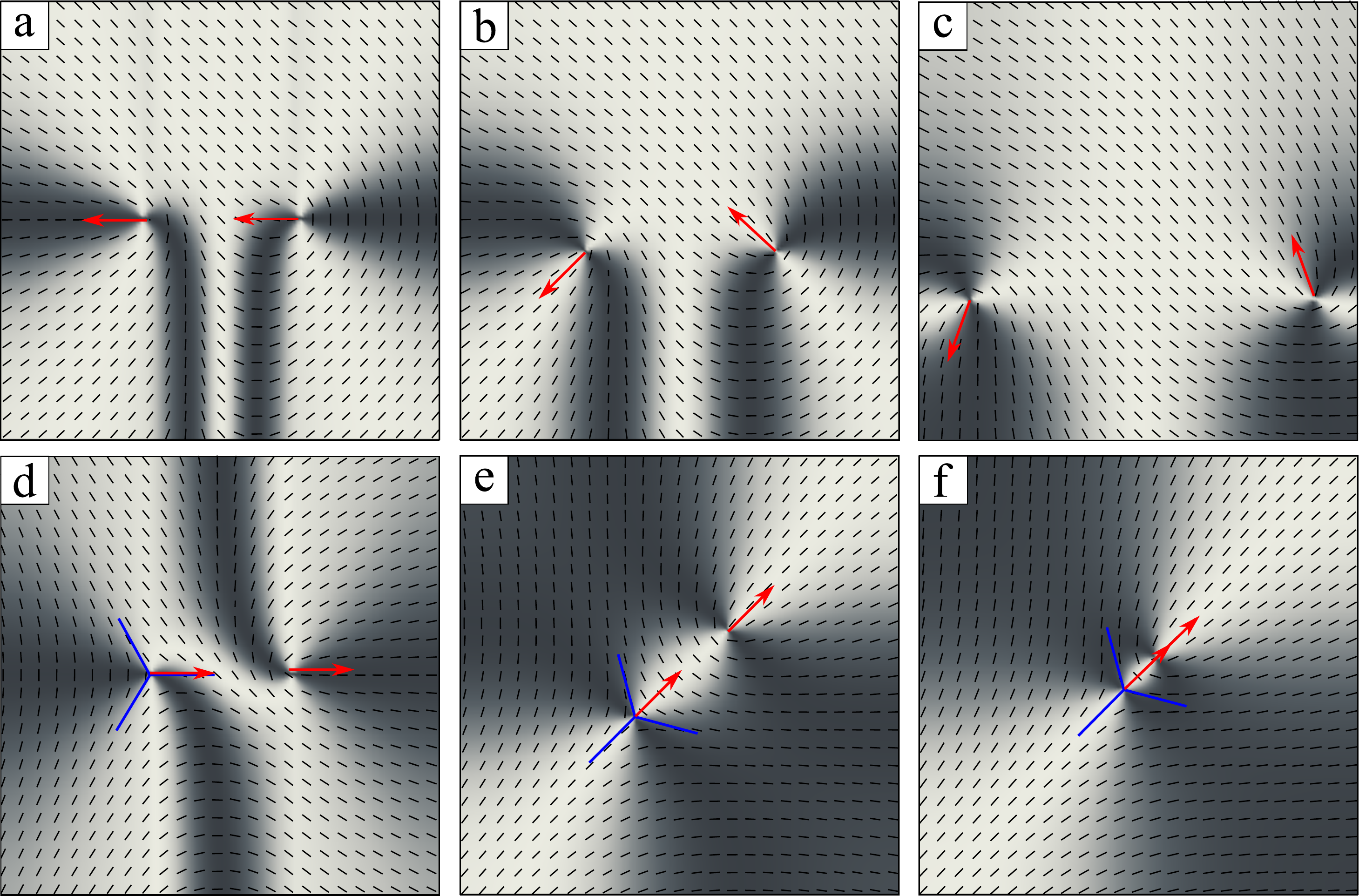}  
\caption{\label{fig:dynamics}The temporal evolution of a pair of $+1/2$ (a-c) and $\pm 1/2$ (d-f) disclinations obtained from a numerical integration of Eq. \eqref{eq:dynamics} with $\epsilon/L=5\times 10^{-3}$, initial distance $d/L=1/3$ and initial tilt $\Delta\psi=0$.}
\end{figure}

If the disclinations are now left free to relax, we expect them to rotate and progressively reach the energy-minimizing antiparallel configuration, thus effectively exerting on each other an elastic torque $T=-dE/d\psi$. In order to test this scenario, we have numerically integrated the equation governing the relaxational dynamics of the nematic phase: 
\begin{equation}\label{eq:dynamics}
\frac{dQ_{ij}}{dt} = -\frac{1}{\gamma}\,\frac{\delta E_{\rm LdG}}{\delta Q_{ij}} = \frac{K}{\gamma}\left[\Delta Q_{ij}+\frac{1-S^{2}}{\epsilon^{2}}\,Q_{\ij}\right]\;,
\end{equation}
with $\gamma$ the rotational viscosity \cite{DeGennes:1993}. Eq. \eqref{eq:dynamics} has been integrated using finite differences on a $256\times 256$ grid and Neumann boundary conditions: i.e. $\partial_{\perp}Q_{ij}=0$ at $(x,y)=(\pm L/2,\pm L/2)$, with $\partial_{\perp}$ the normal derivative (see \href{http://wwwhome.lorentz.leidenuniv.nl/~giomi/sup_mat/20150720}{Supplementary Movies S2-4}). Choosing a fixed orientation at the boundary or periodic boundary conditions would restrict the type of defect pair we could consider, as a consequence of the Poincar\'e-Hopf theorem \cite{Kamien:2002}. \revision{Neumann boundary conditions reflects the experimental scenario of weak or no anchoring. Within Landau-de Gennes theory, surface anchoring is often described via the Nobili-Durand energy density \cite{Nobili:1992} given by: $e_{\rm ND}=W \tr(\bm{Q}-\bm{Q}_{0})^{2}$, where $W$ is a constant and $\bm{Q}_{0}$ embodies the preferred orientation at the boundary (typically tangential or orthogonal). Neumann boundary condition corresponds to the case in which $W\sim 0$ and the director at the boundary is free to rotate.}

Fig. \ref{fig:trajectories}a shows the dot product $\bm{p}_{1}\cdot\bm{p}_{2}$ as a function of time for two $+1/2$ disclinations with various initial tilt. Regardless of the initial orientation, the defects relax toward the antiparallel configuration $\bm{p}_{1}=-\bm{p}_{2}$. The relaxation rate is not constant and depends on the distance from the boundary, which becomes progressively shorter as the defects repel. For short times, while the defects are still far from the boundary, the angular displacement decays exponentially as one would expect from energetic considerations. 

The angular dynamics is strongly coupled with the translational dynamics of the core. The latter has been described by Kawasaki \cite{Kawasaki:1984} and Denniston \cite{Denniston:1996} and relies on the decomposition $\theta=\theta_{1}+\theta_{\rm ext}$, where $\theta_{1}=k_{1}\arctan[(y-y_{1})/(x-x_{1})]$ describes the behavior near the core at $\bm{r}_{1}=(x_{1},y_{1})$ and $\theta_{\rm ext}$ describes the departure from the optimal defective configuration and plays the role of an external field. If the core radius is sufficiently small, the dynamics of the core is governed by the equation:
\begin{equation}\label{eq:denniston}
\zeta\frac{d\bm{r}_{1}}{dt} = -2\pi k_{1} K \nabla_{\perp}\theta_{\rm ext}(\bm{r}_{1})\;, 
\end{equation}
with $\nabla_{\perp}=(-\partial_{y},\partial_{x})$ \cite{Denniston:1996}. \revision{The quantity $\zeta$, is an effective drag coefficient proportional to the rotational viscosity $\gamma$ in Eq. \eqref{eq:dynamics} and generally dependent on the core velocity: $\zeta \approx \pi \gamma k^{2} \log(3.6/\mathrm{Er})$, with $\mathrm{Er}=\gamma a |d{\bm r}_{1}/dt|/K$ is the core Eriksen number. In first approximation $\log(3.6/\mathrm{Er})\approx 1$ as a defect typically moves by a few core radii within the nematic relaxational time scale $\tau=\gamma a^{2}/K$, hence $|d{\bm r}_{1}/dt| \approx a/\tau$.}

In the presence of a second defect located at $\bm{r}_{2}=(x_{2},y_{2})$, approximating $\theta_{\rm ext}=k_{2}\arctan[(y-y_{2})/(x-x_{2})]$ and using \eqref{eq:denniston}, yields $\nabla_{\perp}\theta_{\rm ext}(\bm{r}_{1})=-k_{2}\bm{r}_{12}/|\bm{r}_{12}|^{2}$, with $\bm{r}_{12}=\bm{r}_{1}-\bm{r}_{2}$. Thus pairs of disclinations attract or repel each other as over-damped charged particles subject to the two-dimensional Coulomb force $\bm{F}_{12}=-\bm{F}_{21}=2\pi k_{1}k_{2} K \bm{r}_{12}/|\bm{r}_{12}|^{2}$. Fig. \ref{fig:trajectories}b shows the trajectories of the two defects in the central region of the square domain. The additional distortion due to the departure from the anti-parallel configuration introduces a transverse component in $\nabla_{\perp}\theta_{\rm ext}$, causing the defect to follow a curved trajectory (Fig. \ref{fig:dynamics}a-c). Everything we discussed for pairs of $+1/2$ disclinations holds for pairs of $-1/2$ disclinations as well since the energy \eqref{eq:landau_de_gennes} is invariant under the mirror-reflection (i.e. $Q_{xy}\rightarrow-Q_{xy}$) that transforms $+1/2$ into $-1/2$ defects. 

The case of a $\pm 1/2$ pair appears, on the other hand, very different as demonstrated by the dynamics of the product $\bm{p}_{1}'\cdot\bm{p}_{2}$ shown Fig. \ref{fig:trajectories}c. In this case, the defects maintain the same orientation during the entire annihilation process, while the trajectories of the cores reveal the same prominent angular dependence observed for the $+1/2$ pair (Fig. \ref{fig:trajectories}d and Fig. \ref{fig:dynamics}d-f).

\subsection{Orientational dynamics of like-sign disclination on an infinite plane}

\begin{figure}[t]
\centering
\includegraphics[width=1\columnwidth]{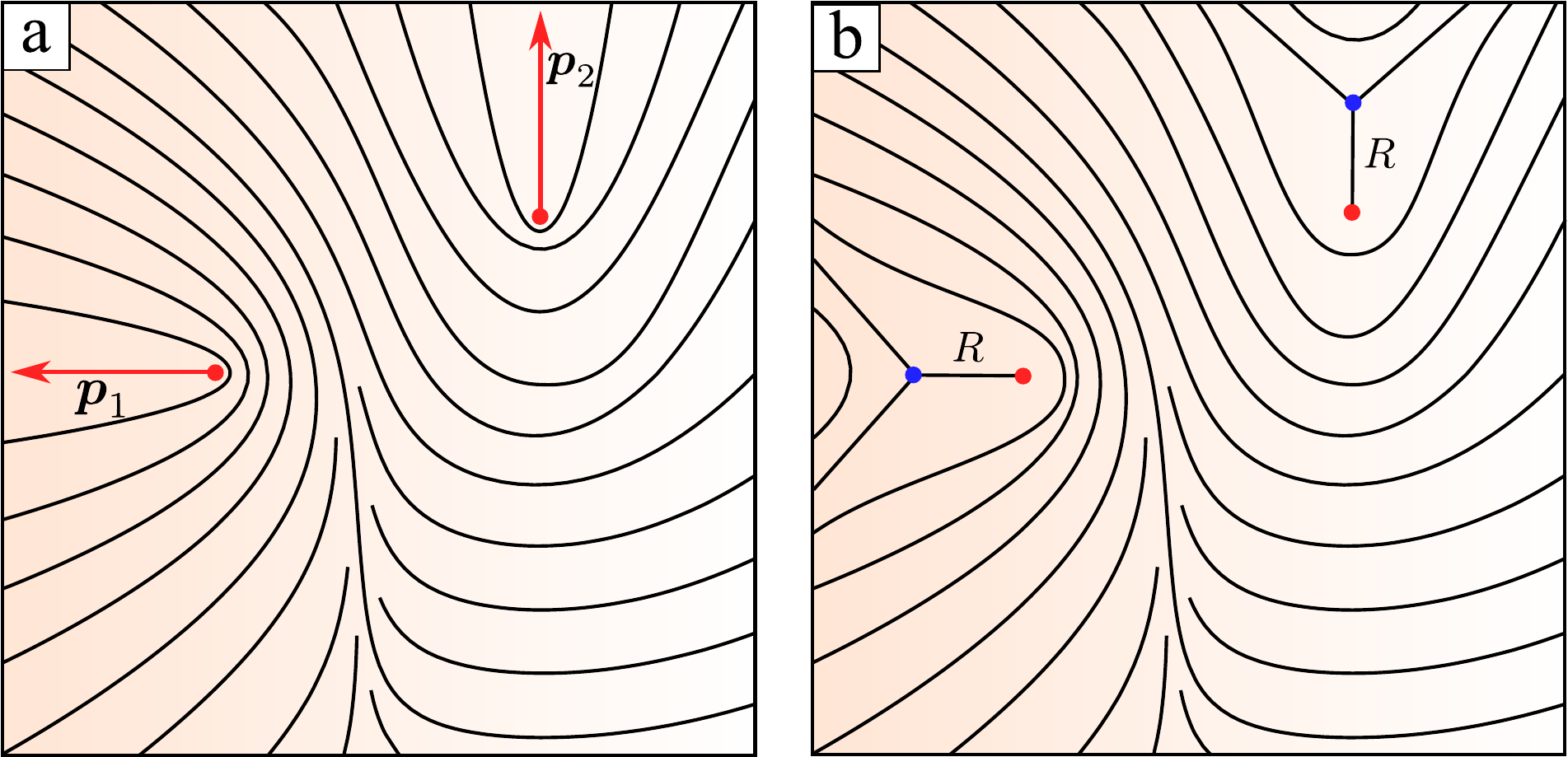}  
\caption{\label{fig:dipole}Illustration of the method used for the derivation of Eq. \eqref{eq:asymptotic}. In order to introduce the energetic contribution due to the relative orientation of two like-sign defects, a pair of $+1/2$ disclinations (a) is replaced by a pair of $\pm 1/2$ dipoles (b), such that $\bm{r}_{i}^{*}=\bm{r}_{i}+R\bm{p}_{i}$ ($i=1,\,2$), with $\bm{r}_{i}$ ($\bm{r}_{i}^{*}$) the position of the positive (negative) defect and $R$ an arbitrary distance. The energy of the original configuration can be recovered by taking $R\rightarrow\infty$.}
\end{figure}

From the previous section it should be clear that a pure particle description of the rotational dynamics of $\pm 1/2$ disclinations is not possible because the energy of a given configuration, hence the torque, depends on the structure of the \revision{transition region between the defects}. A special situation, where this limitation does not occur, is represented by the case of two like-sign disclinations, located on an infinite plane at $\bm{r}_{1}$ and $\bm{r}_{2}$ and oriented along the directions $\bm{p}_{1}$ and $\bm{p}_{2}$. This problem, which was preliminarily discussed in Ref. \cite{Keber:2014}, can be addressed by introducing an oppositely charged ``image defect'' at $\bm{r}_{i}^{*}=\bm{r}_{i}+R{\bm p}_{i}$, with $i=1,\,2$ and $R$ an arbitrary distance (Fig. \ref{fig:dipole}). The energy of this auxiliary configuration is given by: 
\begin{equation}
E=-2\pi K\sum_{i<j}^{1,\,4}k_{i}k_{j}\log\frac{|\bm{r}_{i}-\bm{r}_{j}|}{a}\;,
\end{equation}
with $a$ the defect core radius. The energy of the original configuration can be then recovered by taking $R\rightarrow\infty$ while preserving the orientations. The yields, after simple algebraic manipulations:
\begin{equation}\label{eq:asymptotic}
E = E_{\rm self}-2\pi k^{2} K \left[\log\frac{|\bm{r}_{1}-\bm{r}_{2}|}{a}+\frac{1}{2}\log(1-\bm{p}_{1}\cdot\bm{p}_{2})\right]\;,	
\end{equation}
where $E_{\rm self} \sim \log(R/a)$ is a position and orientation independent self-energy. The orientational contribution is then minimal when $\bm{p}_{1}=-\bm{p}_{2}$, while it diverges for $\bm{p}_{1}=\bm{p}_{2}$. More remarkably, the second term in Eq. \eqref{eq:asymptotic} does not depend on the distance between the defects. This latter property, which is reminiscent of the behavior of dislocations in two-dimensional solids \cite{Nabarro:1987,Eisenmann:2005} is due to the fact that, in this construction, the \revision{transition region} between the two disclinations occupies the entire plane. Thus, while the positional interaction still depends on the defect separation, the orientational interaction is delocalized over the entire plane. The torque associated with Eq. \eqref{eq:asymptotic} can be readily calculated:
\begin{equation}\label{eq:torque}
T_{12} = -T_{21}= -\pi k^{2} K\,\frac{\bm{p}_{1}^{\perp}\cdot\bm{p}_{2}}{1-\bm{p}_{1}\cdot\bm{p}_{2}}\;,  
\end{equation}
where $\bm{p}_{i}^{\perp}=(-p_{y},p_{x})$. Consistently with Eqs. \eqref{eq:dynamics} and \eqref{eq:denniston}, we postulate the orientational dynamics resulting from the torque \eqref{eq:torque} to be purely overdamped, so that:
\begin{equation}\label{eq:dpsidt}
\zeta_{\rm r}\frac{d\psi_{i}}{dt} = \pi k^{2} K \sum_{i<j}\cot\left(\frac{\psi_{i}-\psi_{j}}{2}\right)\;.
\end{equation}
with $\zeta_{\rm r}$ an effective rotational drag coefficient. Using Eq. \eqref{eq:dpsidt} one straightforwardly finds:
\begin{equation}
\cos\Delta\psi(t)+1=[\cos\Delta\psi(0)+1]\,e^{-\frac{t}{\tau}}\;,
\end{equation}
with $\tau=\zeta_{\rm r}/(2\pi K k^{2})$. Thus the defects exponentially relax toward the anti-parallel configuration, consistently with the early times dynamics observed in our simulations. The construction out-lined above cannot be extended to the case of a pair of $\pm 1/2$ disclinations as a consequence of the non-uniqueness of the $\bm{p}'$ vector. 

\section{Discussion and conclusion} 
 
Although discovered in the context of active nematics \cite{Keber:2014,DeCamp:2015}, the polar structure of half-strength disclinations is a general property of nematic defects in both passive and active systems. In this article we 
demonstrated that a notion of polarity can be introduced for both $+1/2$ and $-1/2$ disclinations via the the vector $\bm{p}$ defined in Eq. \eqref{eq:p}. Due to the discrete rotational symmetry of half-strength defects, $\bm{p}$ is defined up to rotations by $\pi/(1-k)$, with $k$ the defect turning number. Thus, for $+1/2$ disclinations, $\bm{p}$ spans the entire unit circle, while in the case of $-1/2$ disclinations it is defined up to rotations by $2\pi/3$ (Fig. \ref{fig:snapshots}). As in the case of dislocations in solids, the elastic energy of a pair of like-sign disclinations, depends on their relative orientation, with the antiparallel configuration representing the lowest energy alignment (Fig. \ref{fig:energy}). As a consequence, half-strength disclinations effectively exert torques on each other and, if left free to move, exponentially relax toward the antiparallel configuration (Figs. \ref{fig:trajectories} and \ref{fig:dynamics}). 

While in this article we have laid down the first fundamental concepts of the orientational mechanics of nematic disclinations, much is still to be understood. Both in passive and active nematics, the strong distortion associated with defects fuels hydrodynamic flow, which in turn causes a reorientation of the nematic director \cite{Toth:2002,Giomi:2014,Giomi:2013,Pismen:2013}. The structure of such a \emph{backflow}, is determined by the geometry of the defect, thus by its polarity. This suggests that, in addition to the elastic torques described here, half-strength disclinations can additionally exert hydrodynamic torques. These are expected to lead to a richer and more complex type of orientational interactions, especially in the case of active nematics where backflow effects are more pronounced. 

We acknowledge Zvonimir Dogic, Vincenzo Vitelli and Jonathan Selinger for useful discussions. This work is supported by The Netherlands Organization for Scientific Research (NWO/OCW).

\end{document}